\documentclass{elsart}
\usepackage{amsmath,amssymb}

\newcommand{\cN}{\mathcal{N}}
\newcommand{\cP}{\mathcal{P}}
\newcommand{\cT}{\mathcal{T}}
\newcommand{\cW}{\mathcal{W}}
\newcommand{\rme}{\mathrm{e}}
\newcommand{\rmd}{\mathrm{d}}

\begin{document}

\begin{frontmatter}

\title{A generalized non-Hermitian oscillator Hamiltonian,
 $\cN$-fold supersymmetry and position-dependent mass models}
\author{
Bijan Bagchi}
\ead{bbagchi123@rediffmail.com}
\address{Department of Applied Mathematics, University of Calcutta,
 92 Acharya Prafulla Chandra Road, Kolkata 700009, India}
\author{Toshiaki Tanaka}
\ead{ttanaka@mail.ncku.edu.tw}
\address{Department of Physics, National Cheng Kung University,\\
 Tainan 701, Taiwan, R.O.C.}
\address{National Center for Theoretical Sciences, Taiwan, R.O.C.}


\begin{abstract}
A generalized non-Hermitian oscillator Hamiltonian is proposed that
consists of additional linear terms which break $\cP\cT$-symmetry
explicitly. The model is put into an equivalent Hermitian form by
means of a similarity transformation and the criterion of
$\cN$-fold supersymmetry with a
position-dependent mass is shown to reside in it.
\end{abstract}


\end{frontmatter}

\section{Introduction}

The study of non-Hermitian quantum systems has been a subject
of considerable interest in recent times. In particular, Swanson's
scheme~\cite{Sw04} of a generalized oscillator with the underlying
Hamiltonian expressed in terms of the usual harmonic oscillator
creation and annihilation operators $\eta^{\dag}$ and $\eta$,
namely $H=\omega(\eta^{\dag} \eta+\frac{1}{2})+\alpha
\eta^{2}+\beta(\eta^{\dag})^{2}$ with $\omega,\alpha,\beta$
three real parameters such that $\alpha\neq\beta$ and
$\Omega^{2}=\omega^{2}-4\alpha\beta>0$ has found much
attention~\cite{Jo05,MGH07,BQR05,Qu07c} in the literature in view
of its implicit non-Hermiticity. In fact $H$ is Hermitian only for
the restricted case of $\alpha=\beta$ but non-Hermitian otherwise.
Actually $H$ is $\cP\cT$-symmetric
as is readily demonstrated by applying the trasformation
properties of $\cP:\,\eta\rightarrow-\eta$ and
$\cT:\,\eta\rightarrow\eta$ for all values of $\alpha$ and
$\beta$. Swanson's Hamiltonian is known to possess
a real, positive and discrete spectrum in line with the conjecture of
Bender and Boettcher~\cite{BB98a}.

Sometime ago, Jones~\cite{Jo05} constructed a similarity transformation
to point out that the above Hamiltonian admits of an equivalent
Hermitian representation. Subsequently Musumbu \textit{et al.}~\cite{MGH07}
showed by means of Bogoliubov transformations that a family of
positive-definite metric operators exists for each of which the
corresponding Hermitian counterpart could be worked out.

On the other hand, Bagchi \textit{et al.} exploited~\cite{BQR05} a hidden
symmetry structure to expose the pseudo-Hermitian character~\cite{Mo02a,%
Mo02b,Mo02c,SGH92,KS04} of $H$ that allows access to a generalized
quantum condition. This was followed up by Quesne~\cite{Qu07c,Qu08}
to take up an $su(1,1)$ embedding of $H$ and to obtain a family of
positive definite metrics that facilitate seeking quasi-Hermitian
supersymmetric (SUSY) theories (for references on SUSY quantum
mechanics, see e.g.~\cite{Ju96,Ba00,CKS01}).

A generalized quantum condition enables one~\cite{BQR05} to connect
those physical systems which are describable
by a position dependence in mass by suitably representing the operator
$\eta$. In this note, we show that such position dependent mass (PDM)
models (for the recent development in PDM models, see
e.g.~\cite{Qu07a,Qu07b,MM08} and the references cited therein)
are necessarily endowed with a type A $\cN$-fold
SUSY~\cite{AST01a,Ta03a,Ta06a} structure. As is well known,
the latter is characterized by the anticommutators of fermionic
operators which are polynomials of degree (at most) $\cN$ in bosonic
operators. The framework of $\cN$-fold SUSY (for the general aspects,
see~\cite{AIS93,AST01b,AS03} and the references cited therein) has
proved to be powerful to deal with one body quantum mechanical
system admitting analytical solutions.

\section{$\cN$-fold SUSY in Generalized Swanson's Models}

To keep our discussion as general as possible, we address an extended
Swanson's model defined by
\begin{align}
H=\omega\left(\eta^{\dag}\eta+\frac{1}{2}\right)+\alpha\eta^{2}
 +\beta(\eta^{\dag})^{2}+\gamma \eta+\delta\eta^{\dag},
\end{align}
where $\omega,\alpha,\beta,\gamma,\delta$ are real parameters.
It differs from the original Swanson's model in the presence of the
linear terms which break $\cP\cT$-symmetry explicitly.

Let us adopt for $\eta$ the most general first-order differential
operator, namely
\begin{align}
\eta=a(x)\frac{\rmd}{\rmd x}+b(x),
\end{align}
where $a(x)$ and $b(x)$ are arbitrary functions. As a result $H$
becomes
\begin{align}
H=-\tilde{\omega}\frac{\rmd}{\rmd x}{a(x)}^{2}\frac{\rmd}{\rmd x}
 +b_{1}(x)\frac{\rmd}{\rmd x}+c_{2}(x),
\end{align}
where  $\tilde{\omega}=\omega-\alpha-\beta$, and the functions
$b_{1}(x)$ and $c_{2}(x)$ stand for
\begin{align}
b_{1}(x)=&\,(\alpha-\beta)a(x)\left(2b(x)-a'(x)\right)
 +(\gamma-\delta)a(x),\\[8pt]
c_{2}(x)=&\,(\tilde{\omega}+2\alpha+2\beta)b(x)^{2}-(\tilde{\omega}
 +\alpha+3\beta)a'(x)b(x)-(\tilde{\omega}+2\beta)a(x)b'(x)
 \nonumber\\
&\,+\beta\left(a(x)a''(x)+a'(x)^{2}\right)+(\gamma+\delta)b(x)
 -\delta a'(x)+\frac{\tilde{\omega}+\alpha+\beta}{2}.
\end{align}
Following a standard procedure, the non-Hermitian operator
$H$ can be transformed into an equivalent Hermitian form by
means of the similarity transformation
\begin{align}
h=\rho H\rho^{-1}=-\tilde{\omega}\frac{\rmd}{\rmd x}a(x)^{2}
 \frac{\rmd}{\rmd x}+V_{\mathrm{eff}}(x),
\label{eq:HeqH}
\end{align}
with the mapping function $\rho$ given by
\begin{align}
\rho=\exp\left(-\frac{1}{2\tilde{\omega}}\int\rmd x\,
 \frac{b_{1}(x)}{a(x)^{2}}\right).
\end{align}
In Eq.~(\ref{eq:HeqH}), the effective potential in terms of $a(x)$
and $b(x)$ reads
\begin{align}
\lefteqn{
V_{\mathrm{eff}}(x)=\left(\frac{(\alpha-\beta)^{2}}{\tilde{\omega}}
 +\tilde{\omega}+2\alpha+2\beta\right)b(x)\left(b(x)-a'(x)\right)
 }\nonumber\\
&-(\tilde{\omega}+\alpha+\beta)a(x)b'(x)
 +\frac{\alpha+\beta}{2}a(x)a''(x)+\frac{1}{4}
 \left(\frac{(\alpha-\beta)^{2}}{\tilde{\omega}}
 +2\alpha+2\beta\right)a'(x)^{2}\nonumber\\
&+\left(\frac{(\alpha-\beta)(\gamma-\delta)}{\tilde{\omega}}
 +\gamma+\delta\right)\left(b(x)-\frac{a'(x)}{2}\right)
 +\frac{(\gamma-\delta)^{2}}{4\tilde{\omega}}
 +\frac{\tilde{\omega}+\alpha+\beta}{2}.
\label{eq:Veff}
\end{align}
Let us consider the case when the commutator of $\eta$ and
$\eta^{\dag}$ is a constant, that is,
\begin{align}
[\eta,\eta^{\dag}]=2a(x)b'(x)-a(x)a''(x)=1.
\label{eq:cometa}
\end{align}
This is equivalent to the relation
\begin{align}
b(x)=\int\frac{\rmd x}{2a(x)}+\frac{a'(x)}{2}.
\label{eq:bofx}
\end{align}
Then, the effective potential (\ref{eq:Veff}) is expressed solely in
terms of $a(x)$ as
\begin{align}
V_{\mathrm{eff}}(x)=a_{1}\left(\int\frac{\rmd x}{2a(x)}\right)^{2}
 +a_{2}\int\frac{\rmd x}{2a(x)}-\frac{\tilde{\omega}}{4}
 \left(2a(x)a''(x)+a'(x)^{2}\right)+\lambda,
\end{align}
where $a_{1}$, $a_{2}$, and $\lambda$ denote the following constants:
\begin{align}
a_{1}&=\frac{(\alpha-\beta)^{2}}{\tilde{\omega}}+\tilde{\omega}
 +2\alpha+2\beta,
\label{eq:a1}\\
a_{2}&=\frac{(\alpha-\beta)(\gamma-\delta)}{\tilde{\omega}}
 +\gamma+\delta,\\
\lambda&=\frac{(\gamma-\delta)^{2}}{4\tilde{\omega}}.
\end{align}
The connection to PDM systems can now be identified by applying 
the following set of transformations
\begin{align}
a(x)=\frac{1}{\sqrt{2\tilde{\omega} m(x)}},\qquad
 u(x)=\int\rmd x\sqrt{m(x)}.
\label{eq:aandu}
\end{align}
Written in terms of the equivalent Hermitian Hamiltonian $h$ gives:
\begin{align}
h=-\frac{\rmd}{\rmd x}\frac{1}{2m(x)}\frac{\rmd}{\rmd x}
 +\frac{\tilde{\omega}}{2}a_{1}u(x)^{2}
 +\sqrt{\frac{\tilde{\omega}}{2}}a_{2}u(x)+\frac{m''(x)}{8m(x)^{2}}
 -\frac{7m'(x)^{2}}{32m(x)^{3}}+\lambda.
\label{eq:HeqH2}
\end{align}
It readily follows from (\ref{eq:HeqH2}) that $h$ has $\cN$-fold
SUSY. More precisely, $h$ belongs to type A $\cN$-fold SUSY PDM
Hamiltonian, case I, discussed in~\cite{Ta06a}.

\section{Further Generalization}

What happens if the commutator of $\eta$ and $\eta^{\dag}$ is
not a constant? To observe it, consider the general form of type A
$\cN$-fold SUSY PDM Hamiltonians given by~\cite{Ta06a}
\begin{align}
H_{\cN}^{\pm}=-\frac{\rmd}{\rmd x}\frac{1}{2m(x)}\frac{\rmd}{\rmd x}
 +V_{\cN}^{\pm}(u)+\frac{m''(x)}{8m(x)^{2}}-\frac{7m'(x)^{2}}{32m(x)^{3}},
\label{eq:NfPDM}
\end{align}
where
\begin{align}
V_{\cN}^{\pm}=&\,\frac{Q(z)^{2}}{2f'(u)^{2}}-\frac{{\cN}^{\,2}-1}{24}
 \left(\frac{2f'''(u)}{f'(u)}-\frac{3f''(u)^{2}}{f'(u)^{2}}\right)\nonumber\\
&\,\pm\left.\frac{\cN}{2}\left(\frac{f''(u)}{f'(u)^{2}}Q(z)-Q'(z)\right)
 -R\right|_{z=f(u)}.
\label{eq:Nfpot}
\end{align}
In (\ref{eq:Nfpot}), $R$ is a constant, $Q(z)$ is a polynomial of (at most)
second degree in $z$, and $f(u)$ is one of the following complex functions
which characterize the different cases of type A $\cN$-fold SUSY
(cf. Section~6 and Table~1 in~\cite{Ta06a}):
\begin{align}
u,\qquad u^{2},\qquad \rme^{2\sqrt{\nu}u},\qquad \cosh2\sqrt{\nu}u,
 \qquad\wp(u).
\label{eq:class}
\end{align}
Using (\ref{eq:aandu}), the type A $\cN$-fold SUSY PDM Hamiltonian
(\ref{eq:NfPDM}) acquires a form resembling a transformed oscillator
model as follows:
\begin{align}
H_{\cN}^{\pm}=-\tilde{\omega}\frac{\rmd}{\rmd x}a(x)^{2}
 \frac{\rmd}{\rmd x}+V_{\cN}^{\pm}(u)-\frac{\tilde{\omega}}{4}
 \left(2a(x)a''(x)+a'(x)^{2}\right).
\label{eq:NfPDM2}
\end{align}
Comparing (\ref{eq:NfPDM2}) with (\ref{eq:HeqH}), we immediately see that
the operator $h$ is of type A $\cN$-fold SUSY if only $b(x)$ is a function
of $a(x)$ and $u(x)$ such that $V_{\mathrm{eff}}$ depends on $a(x)$
just as the last term in (\ref{eq:NfPDM2}). This is shown to be possible if
and only if $b(x)$ is assigned the following form:
\begin{align}
b(x)=B_{0}(u)+B_{2}(u)a'(x),
\label{eq:bofx2}
\end{align}
where $B_{0}(u)$ and $B_{2}(u)$ are for the moment arbitrary functions.
Indeed substituting (\ref{eq:bofx2}) in (\ref{eq:Veff}), one obtains
\begin{align}
V_{\mathrm{eff}}(x)=F_{0}(u)+F_{1}(u)a'(x)+F_{2}(u)a(x)a''(x)
 +F_{3}(u)a'(x)^{2},
\end{align}
where
\begin{align}
F_{0}(u)&=a_{1}B_{0}(u)^{2}+a_{2}B_{0}(u)
 -\frac{\tilde{\omega}+\alpha+\beta}{\sqrt{2\tilde{\omega}}}B'_{0}(u)
 +\lambda+\frac{\tilde{\omega}+\alpha+\beta}{2},\\
F_{1}(u)&=\left(a_{1}B_{0}(u)+\frac{a_{2}}{2}\right)\left(2B_{2}(u)-1\right)
 -\frac{\tilde{\omega}+\alpha+\beta}{\sqrt{2\tilde{\omega}}}B'_{2}(u),\\
F_{2}(u)&=-(\tilde{\omega}+\alpha+\beta)B_{2}(u)
 +\frac{\alpha+\beta}{2},\\
F_{3}(u)&=a_{1}\left(B_{2}(u)^{2}-B_{2}(u)+\frac{1}{4}\right)
 -\frac{\tilde{\omega}}{4}.
\end{align}
We are thus led to a necessary condition for $h$ to be of
type A $\cN$-fold SUSY summarized by the following set of
equations:
\begin{align}
F_{1}(u)=0,\qquad 2F_{2}(u)=4F_{3}(u)=-\tilde{\omega}.
\label{eq:Nfcond}
\end{align}
It is straightforward to observe that the non-trivial solution to
(\ref{eq:Nfcond}) is
\begin{align}
B_{2}(u)=\frac{1}{2},
\end{align}
and we have the desired form
\begin{align}
b(x)=B_{0}(u)+\frac{a'(x)}{2}.
\label{eq:bofx3}
\end{align}
In this case, the commutator of $\eta$ and its transposition is not
a constant but can be an arbitrary function of $u(x)$:
\begin{align}
[\eta,\eta^{\dag}]=2a(x)b'(x)-a(x)a''(x)=\sqrt{\frac{2}{\tilde{\omega}}}
 B'_{0}(u).
\end{align}
If we put $B_{0}(u)=\sqrt{\tilde{\omega}/2}\,u$, we reproduce
all the previous results, namely, (\ref{eq:cometa})--(\ref{eq:HeqH2}).
With the condition (\ref{eq:bofx3}) satisfied, the equivalent Hermitian
Hamiltonian operator $h$ now becomes
\begin{align}
h=-\tilde{\omega}\frac{\rmd}{\rmd x}a(x)^{2}\frac{\rmd}{\rmd x}
 +F_{0}(u)-\frac{\tilde{\omega}}{4}\left(2a(x)a''(x)+a'(x)^{2}\right).
\end{align}
Finally, comparing with (\ref{eq:NfPDM2}), we obtain the necessary and
sufficient condition for the operator $h$ to be type A $\cN$-fold SUSY
as
\begin{align}
F_{0}(u)=a_{1}B_{0}(u)^{2}+a_{2}B_{0}(u)-a_{3}B'_{0}(u)
 +a_{4}=V_{\cN}^{\pm}(u),
\label{eq:F0}
\end{align}
where $a_{3}$ and $a_{4}$ are given by
\begin{align}
a_{3}=-\frac{\tilde{\omega}+\alpha+\beta}{\sqrt{2\tilde{\omega}}},
 \qquad a_{4}=\lambda+\frac{\tilde{\omega}+\alpha+\beta}{2}.
\label{eq:a3a4}
\end{align}
To derive a solution to Eq.~(\ref{eq:F0}), it is convenient to convert
it to a standard form of Riccati equation
\begin{align}
\phi'(u)=a_{1}a_{3}^{-2}\phi(u)^{2}+a_{2}a_{3}^{-1}\phi(u)
 -V_{\cN}^{\pm}(u)+a_{4},
\label{eq:phi}
\end{align}
by the substitution $B_{0}(u)=a_{3}^{-1}\phi(u)$. Linearizing
(\ref{eq:phi}) with the help of introducing $\phi(u)=-a_{3}^{2}
\psi'(u)/a_{1}\psi(u)$ gives
\begin{align}
-\psi''(u)+\frac{a_{2}}{a_{3}}\psi'(u)+\frac{a_{1}}{a_{3}^{2}}
 \left(V_{\cN}^{\pm}(u)-a_{4}\right)\psi(u)=0.
\end{align}
Lastly, to eliminate the first derivative term, we make an ansatz
\begin{align}
\psi(u)=\exp\left(\frac{a_{2}}{2a_{3}}u\right)\hat{\psi(u)},
\end{align}
which leads to
\begin{align}
\left[-\frac{1}{2}\frac{\rmd^{2}}{\rmd u^{2}}+\frac{a_{1}}{2a_{3}^{2}}
 V_{\cN}^{\pm}(u)-\frac{a_{1}a_{4}}{2a_{3}^{2}}+\frac{a_{2}^{2}}{8a_{3}^{2}}
 \right]\hat{\psi}(u)=0.
\label{eq:Sch}
\end{align}
Equation (\ref{eq:Sch}) is nothing but a Schr\"{o}dinger equation subject
to the potential $a_{1}V_{\cN}^{\pm}(u)/2a_{3}^{2}$.

An interesting possibility then emerges when $a_{1}=2a_{3}^{2}$, or
equivalently from (\ref{eq:a1}) and (\ref{eq:a3a4}), when
$\alpha\beta=0$.
In this case, the Schr\"{o}dinger operator in (\ref{eq:Sch}) itself
has type A $\cN$-fold SUSY (with the constant mass $m=1$).
Since $\cN$-fold SUSY is essentially equivalent to
quasi-solvability~\cite{AST01b}, the operator is always
quasi-solvable (and is solvable if it does not substantially depend
on $\cN$), which means that we can obtain (at most) $\cN$ analytic
(local) solutions to Eq.~(\ref{eq:Sch}) in a closed form having
the following representation:
\begin{align}
\hat{\psi}(u)=P_{{\cN}-1}(f(u))e^{-{{\cW}_{{\cN}}(u)}},
\label{eq:psihat}
\end{align}
where $P_{n}$ is a polynomial of (at most) $n$th degree,
$\cW_{\cN}$ is a (generalized) superpotential, and $f(u)$
is a function which characterizes one of the cases of type A
$\cN$-fold SUSY (see, Eq.(\ref{eq:class})). From (\ref{eq:Sch})
and (\ref{eq:psihat}), we arrive at the following
specific representation of $B_{0}(u)$:
\begin{align}
B_{0}(u)=-\frac{a_{3}\hat{\psi}'(u)}{a_{1}\hat{\psi}(u)}
 -\frac{a_{2}}{2a_{1}}
 =-\frac{a_{3}}{a_{1}}\left[\frac{f'(u)P'_{{\cN}-1}
 (f(u))}{P_{{\cN}-1}(f(u))}-\cW'_{\cN}(u)\right]
 -\frac{a_{2}}{2a_{1}}.
\label{eq:B0}
\end{align}
Here we note that since the Schr\"{o}dinger equation (\ref{eq:Sch})
is just a differential equation but not an eigenvalue problem,
the normalizability of the solutions including (\ref{eq:psihat}) is
irrelevant.

It is also important to note that in any case of $a_{1}\neq 2a_{3}^{2}$
(or $\alpha\beta\neq 0$) the Schr\"{o}dinger operator in (\ref{eq:Sch})
is in general not quasi-solvable since any change of overall multiplicative
factor of the potential $V_{\cN}^{\pm}\rightarrow\rho V_{\cN}^{\pm}$
breaks quasi-solvability, the fact that is originated from the form
invariance of the quasi-solvable potentials under the scale
transformation $u\rightarrow \rho u$ (cf. Section~4 in \cite{GT05}). 
However, solvability of the operator remains under the transformation
$V_{\cN}^{\pm}\rightarrow\rho V_{\cN}^{\pm}$ since the effect
can be absorbed by the rescaling of the variable $u$ without
violating solvability.
Hence, an analytic solution of (\ref{eq:Sch}) is still obtainable when
the polynomial $Q(z)$ in (\ref{eq:Nfpot}) is at most first-degree in
$z$ for the cases I-IV of type A $\cN$-fold SUSY. Note also that
the formula (\ref{eq:B0}) continues to be valid.
It would be an interesting problem to examine whether we could
obtain the complete list of possible explicit forms of $B_{0}(u)$.

\begin{ack}
One of us (BB) gratefully acknowledges Prof. C.-H. Chen's
 hospitality during his visit at the National Cheng Kung
 University, Taiwan where this work was carried out.
 This work was partially supported by the National Cheng Kung
 University under the grant No. OUA:95-3-2-071 (TT).
\end{ack}


\bibliography{refsels}

\begin{thebibliography}{10}
\expandafter\ifx\csname url\endcsname\relax
  \def\url#1{{\tt #1}}\fi
\expandafter\ifx\csname urlprefix\endcsname\relax\def\urlprefix{URL }\fi
\providecommand{\eprint}[2][]{\url{#2}}

\bibitem{Sw04}
M.~S. Swanson, J. Math. Phys. 45 (2004) 585.

\bibitem{Jo05}
H.~F. Jones, J. Phys. A: Math. Gen. 38 (2005) 1741.
\newblock \eprint{quant-ph/0411171}.

\bibitem{MGH07}
D.~P. Musumbu, H.~B. Geyer, and W.~D. Heiss, J. Phys. A: Math. Theor. 40 (2007)
  F75.
\newblock \eprint{quant-ph/0611150}.

\bibitem{BQR05}
B.~Bagchi, C.~Quesne, and R.~Roychoudhury, J. Phys. A: Math. Gen. 38 (2005)
  L647.
\newblock \eprint{quant-ph/0508073}.

\bibitem{Qu07c}
C.~Quesne, J. Phys. A: Math. Theor. 40 (2007) F745.
\newblock \eprint{0705.2868[math-ph]}.

\bibitem{BB98a}
C.~M. Bender and S.~Boettcher, Phys. Rev. Lett. 80 (1998) 5243.
\newblock \eprint{physics/9712001}.

\bibitem{Mo02a}
A.~Mostafazadeh, J. Math. Phys. 43 (2002) 205.
\newblock \eprint{math-ph/0107001}.

\bibitem{Mo02b}
A.~Mostafazadeh, J. Math. Phys. 43 (2002) 2814.
\newblock \eprint{math-ph/0110016}.

\bibitem{Mo02c}
A.~Mostafazadeh, J. Math. Phys. 43 (2002) 3944.
\newblock \eprint{math-ph/0203005}.

\bibitem{SGH92}
F.~G. Scholtz, H.~B. Geyer, and F.~J.~W. Hahne, Ann. Phys. 213 (1992) 74.

\bibitem{KS04}
R.~Kretschmer and L.~Szymanowski, Phys. Lett. A 325 (2004) 112.
\newblock \eprint{quant-ph/0305123}.

\bibitem{Qu08}
C.~Quesne, J. Phys. A: Math. Theor. 41 (2008) 244022.
\newblock ({P}roceedings), \eprint{0710.2453[math-ph]}.

\bibitem{Ju96}
G.~Junker, Supersymmetric {M}ethods in {Q}uantum and {S}tatistical {P}hysics
  (Springer, Berlin, 1996).

\bibitem{Ba00}
B.~K. Bagchi, Supersymmetry in {Q}uantum and {C}lassical {M}echanics (Chapman
  and Hall/CRC, Florida, 2000).

\bibitem{CKS01}
F.~Cooper, A.~Khare, and U.~Sukhatme, Supersymmetry in {Q}uantum {M}echanics
  (World Scientific, Singapore, 2001).

\bibitem{Qu07a}
C.~Quesne, J. Phys. A: Math. Theor. 40 (2007) 13107.
\newblock \eprint{0705.0862[math-ph]}.

\bibitem{Qu07b}
C.~Quesne, SIGMA 3 (2007) 067.
\newblock \eprint{0705.2577[math-ph]}.

\bibitem{MM08}
O.~Mustafa and S.~H. Mazharimousavi, J. Phys. A: Math. Theor. 41 (2008) 244020.
\newblock ({P}roceedings), \eprint{0707.3738[quant-ph]}.

\bibitem{AST01a}
H.~Aoyama, M.~Sato, and T.~Tanaka, Phys. Lett. B 503 (2001) 423.
\newblock \eprint{quant-ph/0012065}.

\bibitem{Ta03a}
T.~Tanaka, Nucl. Phys. B 662 (2003) 413.
\newblock \eprint{hep-th/0212276}.

\bibitem{Ta06a}
T.~Tanaka, J. Phys. A: Math. Gen. 39 (2006) 219.
\newblock \eprint{quant-ph/0509132}.

\bibitem{AIS93}
A.~A. Andrianov, M.~V. Ioffe, and V.~P. Spiridonov, Phys. Lett. A 174 (1993)
  273.
\newblock \eprint{hep-th/9303005}.

\bibitem{AST01b}
H.~Aoyama, M.~Sato, and T.~Tanaka, Nucl. Phys. B 619 (2001) 105.
\newblock \eprint{quant-ph/0106037}.

\bibitem{AS03}
A.~A. Andrianov and A.~V. Sokolov, Nucl. Phys. B 660 (2003) 25.
\newblock \eprint{hep-th/0301062}.

\bibitem{GT05}
A.~Gonz{\'a}lez-L{\'o}pez and T.~Tanaka, J. Phys. A: Math. Gen. 38 (2005) 5133.
\newblock \eprint{hep-th/0405079}.

\end{thebibliography}
\bibliographystyle{npb}

\end{document}